\documentclass[preprint,electronic]{vgtc}             

\graphicspath{{figures/}{pictures/}{images/}{./}} 

\usepackage{times}                     

\usepackage{tabu}                      
\usepackage{booktabs}                  
\usepackage{lipsum}                    
\usepackage{mwe}                       

\usepackage{mathptmx}                  
\usepackage{mdframed}
\usepackage{balance}

\onlineid{0}

\vgtccategory{Research}

\vgtcinsertpkg

\renewcommand*{\backref}[1]{
}

\title{From Data to Insight: Using Contextual Scenarios to Teach Critical Thinking in Data Visualisation}

\author{Jonathan C. Roberts\thanks{e-mail: j.c.roberts@bangor.ac.uk}\\ %
        \scriptsize Bangor University, UK %
\and Peter Butcher \thanks{e-mail: 
p.butcher@bangor.ac.uk}\\ %
     \scriptsize Bangor University, UK %
     \and Panagiotis D. Ritsos \thanks{e-mail: p.ritsos@bangor.ac.uk}\\ %
     \scriptsize Bangor University, UK %
}

\abstract{
This paper explores the use of scenario-based visualisation examples as a pedagogical strategy for teaching students the complexities of data insight, representation, and interpretation. Teaching data visualisation often involves explaining intricate issues related to data management and the challenges of presenting data meaningfully. In this work, we present a series of data-driven scenarios. These concise stories depict specific situations, and are created to help the educators highlight key concerns in data communication, such as chart selection, temporal versus categorical comparison, visual bias, and narrative framing. By grounding these examples in real-world contexts, students are encouraged to critically assess not only what the data shows, but how and why it is shown that way. The paper presents a collection of example scenarios, that educators can use for their own lessons; the work fits with a larger project on looking at critical thinking in the classroom, and developing appropriate tools.  We also start to abstract principles, from our approach, so that others can develop their own scenarios for their teaching. Our approach aligns with principles of authentic and scenario-based learning, using real-world contexts to foster critical engagement with data.
} 


\preprinttext{\parbox{0.9\textwidth}{$\copyright$ \the\year~IEEE. This is the author's version of the paper for IEEE EduVis 2025. IEEE VIS Workshop on Visualization Education, Literacy, and Activities 2025, Vienna.}}

\begin{document}


\firstsection{Introduction}


\maketitle

Teaching courses in data visualisation, information visualisation, or data science often presents a challenge for the teacher: how to effectively develop students' critical thinking about data. While experienced data visualisation practitioners understand that meaningful visualisations stem from a deep understanding of the data, students frequently lack the analytical skills needed to question, interpret, and visualise data appropriately. In our recent work on the Critical Design Strategy~\cite{Roberts2023CriticalCreative,RobertsAlnjarOwenRitsos2026}, we introduced an approach for evaluating potential visualisation designs, whereas previously we introduced the Critical Thinking Sheet~\cite{RobertsRitsos2020CTS} and the use of explanatory frameworks~\cite{Roberts_ETAL2018_EVF}. However, structuring critical thinking in lectures remains a challenge; when delivering complex issues students can get lost in the detail, or worse may lose interest in the lecture. In particular, foundational issues, such as the origin of the data, who collected it, and under what conditions, can be challenging to explain. There are lots of concerns to raise and discuss in a lecture, and it can be difficult for the teacher to organise it in a clear way. Faced with a list of abstract facts and many considerations to make, students may quickly lose interest. As a result, lectures on these topics can feel less compelling and fail to engage learners effectively. Yet these very issues play a critical role in framing and shaping the final visualisation. Helping students grasp the context and structure of data is not only essential, but transformative: when learners truly understand the data, they are far better equipped to visualise it effectively.

Our solution is to ground abstract concepts in real-world contexts. Our goal is to make interesting yet accessible scenarios, that can be used to hook on the discussion of other issues. The scenarios help to drive the lesson, and provide a structure around which methods and questions (that are often dry and boring to consider) can be made more interesting and fun. Our scenarios are rooted in the principles of authentic learning. Our method involves anchoring each example in a carefully crafted scenario to create a meaningful hook for the lesson. 
Over several years of using this approach, we began to notice recurring patterns in the structure of our scenarios. Each one naturally evolved to include a title and a short descriptive text, which serves as a narrative entry point into the topic. This text can be used to introduce the example and guide the class discussion. 

In this paper, we introduce our scenario-based teaching strategy and present initial examples of its application, and initial structuring and generalisation of the scenarios. Our long-term aim is to develop a suite of reusable scenario-based examples that can be adopted and adapted by other educators and researchers. We  seek to further develop and formalise the framework, with the intention of extending it to address broader challenges in data visualisation. The focus of our current work is specifically on the data itself and how these scenarios can help the teacher discuss the complexities, limitations, and contextual factors that influence how data should be visualised.

\section{Background and design of the scenarios}
\label{sec:background}

Over more than 15 years of teaching visualisation across a range of academic levels, we have consistently encountered a core challenge: helping students engage meaningfully with the data behind the visualisation. While creating visual outputs can be exciting and visually appealing, the deeper issues surrounding the data (its origin, quality, structure, and limitations) are often seen as dry or overly complex. Yet, a strong understanding of these data-related issues is essential for producing accurate, meaningful, and ethical visualisations.

When teaching data visualisation, we often ask students to select their own datasets. However, this freedom can present challenges. Students must consider the suitability of their chosen topic (ideally something positive and constructive), the availability and accessibility of the data, whether it contains sufficient volume, and whether the variability in the data supports meaningful visual analysis. This requires students to think ahead, and anticipate how the data might be used and visualised. Early in our teaching, we introduce the idea of a ``data triage'': a rapid process of exploring online data sources, identifying potential datasets, and evaluating their fitness for purpose (in our case, for their coursework or assessment). To support this process, we teach students to apply critical thinking to both the data and the scenario it represents. We now structure these skills around a set of curated data example scenarios designed to guide their thinking and support deeper analysis.

As a School of Computer Science, we offer programmes in Computer Science, Games Development, and Data Science. Within these, we teach a suite of visual computing modules, including Information Visualisation (part of our taught Masters in Data Science and Advanced Computer Science), Creative Visualisation (for undergraduate computing students), User Experience and Human-Computer Interaction, and Web Technology. Across these modules, we have seen the importance of developing strategies that help students critically engage with data in ways that are both accessible and intellectually stimulating. 

Like many educators, we incorporate examples into our teaching to help explain abstract ideas, often using a picture, diagram, or short description to illustrate a concept. However, we have taken this approach further by actively using example scenarios that go beyond isolated illustrations. Rather than simply showing examples to support a point, we design lectures around a series of worked scenarios. We have pivoted the lecture around one or more scenarios. These scenarios represent real-world vignettes that are authentic in nature. They present situations that students are likely to encounter in professional practice after completing their course. By grounding the teaching in realistic and meaningful contexts, we aim to foster deeper engagement and critical thinking around both the data and its visualisation.

Our development of these rich example scenarios began approximately six years ago, and before the COVID-19 pandemic. At the time, our team was in the process of reorganising the teaching of our visualisation and user interface modules. This led to in-depth discussions about how best to teach data visualisation in a way that was both comprehensive and engaging. We agreed that students needed exposure to key areas such as information visualisation and design, data use and management, user interface development, web technologies, and creativity. Each of these areas addresses a distinct aspect of the visualisation design space, collectively supporting the development of robust data science and visualisation skills. As a result, we structured our curriculum into a set of focused modules, including Information Visualisation, Data Systems and Management, User Experience and HCI, Creative Visualisation, and Web Technologies. Especially for the Information Visualisation and Data Management programmes we wanted to consider authentic lessons and examples. 

Initially, we used standalone examples to illustrate key points, but found that they were disconnected from the core learning goals. Over the past two years, we have shifted to structured scenarios that anchor our lectures and support authentic learning through realistic, meaningful contexts.
By centring our teaching around these scenarios, we can explore them in depth: clarifying the real-world context, identifying the roles of different stakeholders, defining the purpose of the visualisation, and examining a range of viable design options. This opens space for critical discussion, where students are encouraged to compare and critique visualization choices, consider their effectiveness for specific audiences, and evaluate the utility of each design in addressing the objectives of the scenario.

Each data scenario is a short, descriptive piece of writing. They are often brief episodes that capture a particular moment, a scene or ideas. They help people focus on a single impression or an aspect of that narrative without going into too much detail, but each focus on data. It's like a short story or scenario that sets the scene for a specific data challenge or concept, providing just enough background and detail to engage students and stimulate critical thinking. Each are targeted to illustrate key points without overwhelming learners with complexity. They serve as a hook to introduce an example dataset and its visualisation, encouraging students to explore the underlying issues, choices, and insights involved.

\section{Related Work}
\label{sec:related}

The use of examples in teaching visualisation is well established. Classic cases, such as Anscombe’s Quartet~\cite{CorrellETAL2019} for demonstrating statistical nuance, or scatter plots to illustrate clusters to explain machine learning algorithms, offer powerful visual representations that help learners grasp abstract concepts. However, these examples are often presented in isolation, without rich contextual scenarios that connect them to real-world data challenges. An effective way to deepen understanding is by pairing visualisation teaching with structured learning activities: give some theory and then follow it by a practical task~\cite{RobertsETAL2022}. These activities not only help solidify conceptual understanding but also create space for dialogue, reflection, and questioning. Visualisations have also been employed to support the explanation of complex ideas, such as through the use of data comics~\cite{BachETALComics2018} and scenarios have been explored as a means to frame different forms of evaluation in visualisation research~\cite{LamETAL2012}. Together, these approaches highlight the pedagogical value of combining narrative, visual, and interactive elements to support critical thinking in visualisation education.

While narratives can be a way to display data~\cite{segel2010narrative} researchers have also demonstrated that real-world scenarios enhance critical thinking~\cite{Self2017}. Van Meter and Firetto~\cite{VanMeter2024} discussed challenges students face when constructing data visualisations, they emphasise that students require a broad understanding and oversight of their actions while constructing visualisations, and that scenario and activity-based interventions to strengthen critical reflection. Roberts et al.\ \cite{roberts2022reflections} summarise nine strategies, many scenario-driven, that incorporate creative, reflective, and evaluative thinking into visualization education. These scenarios are similar to `thought experiments'. Blythe and Encinas~\cite{BlytheEncinas2018} explore how imaginative narratives, particularly research fiction and design-oriented thought experiments, can function as rigorous methods in Human-Computer Interaction (HCI) and design research. Similarly, Panda~\cite{Panda2024} positions thought experiments as scenario-based narratives, and place strong emphasis on the storytelling aspect of each scenario. In contrast, our approach grounds each example in real-world contexts, incorporating concrete details about the data and its associated visualisations. This enables us to abstract and analyse key scenario components as part of a future framework for visualisation pedagogy.

Scenario-based teaching closely aligns with the principles of authentic assessment, where learning tasks are grounded in realistic, discipline-relevant challenges. Wiggins~\cite{wiggins1993assessing} defines authenticity as designing assessments that mirror the types of tasks professionals face in the real world. This approach promotes meaningful engagement, particularly in data visualisation, where students benefit from working on examples that reflect the complexity and nuance of real-life situations. Rather than simply asking students to create charts from arbitrary datasets, we embed their learning within carefully designed scenarios that reflect actual roles, problems, and decision-making processes. This supports situated learning~\cite{brown1989situated,mcarthur2023rethinking} and allows students to reflect on societal relevance, ethics, and representation, moving beyond technical skill to deeper critical understanding~\cite{DarlingHammondSnyder2000_Authentic}.

To maintain engagement and challenge, our teaching approach integrates personalisation~\cite{bell2010project,kokotsaki2016project}, allowing students to select datasets that align with their own interests and goals~\cite{AlamriETAL2020_PersonalisedLearning}. But to do so, the need to understand the challenges, and have metacognition in their learning. The scenarios used in the lectures help to frame their learning, and give them the structure that they can use and apply to their own dataset. After following a few scenarios in the lectures, students are able to choose their own data, assess its suitability, understand its ethical framing, and start to analyse their own data in the same way.  In this way, the scenarios act as a scaffold to help students think critically not just about visualisation, but about the social and contextual implications of data. Echoing Ashford-Rowe et al.~\cite{ashford2014establishing}, our scenario-based framework encourages challenge, metacognition, real-world application, and regular feedback; core aspects of authentic assessment.

\section{Scenario 1: Oxygen levels change in combustion}
\label{sec:scenario_A}

We acknowledge that this initial scenario is adapted from work of Professor Ken Brodlie, where Brodlie and Asim, conducted research on interpolation techniques, specifically addressing the challenge of preserving positivity when fitting curves~\cite{AsimBrodlie1999}. Their research led to the development of the constrained quadratic Shepard method for interpolation~\cite{Brodlie2005}. We have adapted this example into a simplified scenario for use in our teaching, as described below.

The first scenario is based on a controlled physical experiment. It explores how oxygen levels change in flue gas during the combustion of coal. The dataset records oxygen measurements taken over time, and can be visualised using a line graph or a scatter plot with a fitted curve. While students may not be familiar with the specifics of flue gas or combustion systems, they can generally grasp that this is a real-world physical process. They understand that natural systems tend to change smoothly over time (so abrupt, angular shifts in data are unlikely) and that combustion requires both fuel and oxygen to occur.

\begin{figure}[h]
\begin{mdframed}
    
\begin{minipage}[t]{1.0\columnwidth}
\vspace*{0pt}
\textbf{Monitoring Oxygen Levels in Coal Combustion.} 
In this scenario, a group of Combustion Dynamics researchers built a draft furnace to investigate how oxygen levels change over time, during coal combustion. The group carefully sampled the flue gas using sensors placed in the flue, recording oxygen levels throughout the combustion process. The collected data was recorded in a spreadsheet to enable detailed analysis by the researchers.
\end{minipage}

\vspace{5mm}
\begin{minipage}[t]{0.35\columnwidth}
\vspace*{0pt}
\centering
\begin{tabular}{rr}
\textbf{Time \small{(min)}} & \textbf{Oxygen} \small{\%} \\
\hline
0  & 20.8 \\
2  & 8.8  \\
4  & 4.2  \\
10 & 0.5  \\
28 & 3.9  \\
30 & 6.2  \\
32 & 9.6  \\
\hline
\end{tabular}
\label{tab:oxygen_data}
\end{minipage}
\hfill
\begin{minipage}[t]{0.6\columnwidth}
\vspace*{0pt}
\centering
\includegraphics[alt={Scenario 1. Example focusing on oxygen levels in coal combustion. Figure includes the text description, sample data table and an illustrative diagram of the experiement setup.},width=0.6\columnwidth]{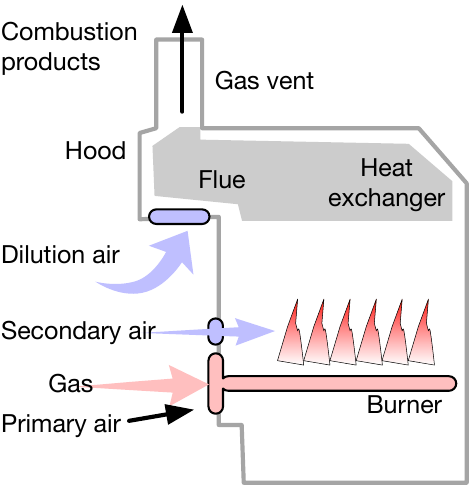}
\label{fig:furnace_plot}
\end{minipage}
\end{mdframed}
\vspace{-5mm}
\caption{Scenario 1. Oxygen concentration in a furnace over time, including a description of the scenario, sample data, and an illustrative diagram. Data from Brodlie et al.\ \cite{Brodlie2005}.}

\end{figure}

\begin{figure}
    \centering
    \includegraphics[alt={Five plots of different visualisations from the monitoring oxygen example.},width=\columnwidth]{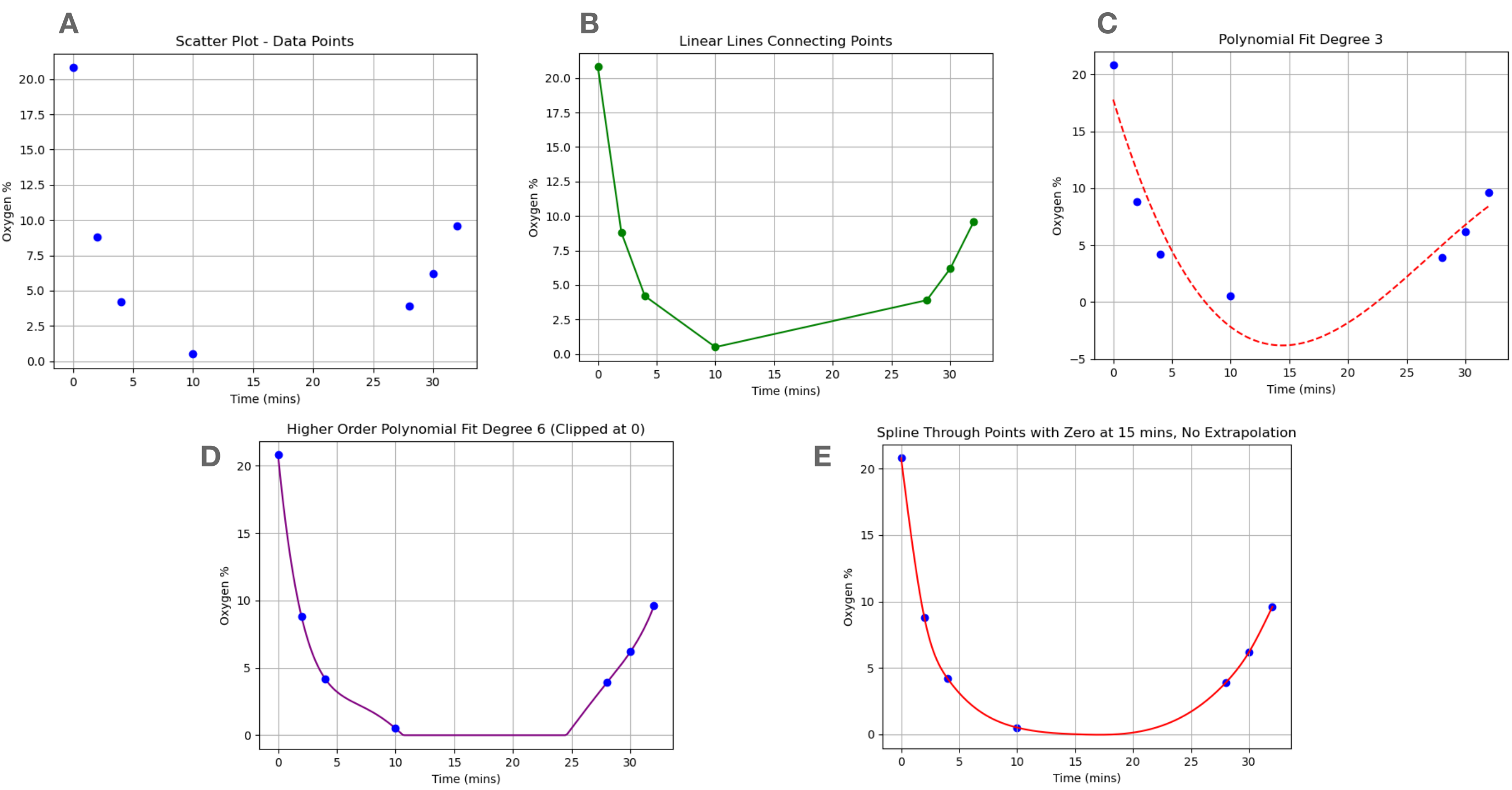}
    \caption{Different ways to plot the data, used to aid with critical thinking; plotted with matplotlib. A) scatter plot, B) linear lines, C) cubic polynomial, D) polynomial fit and clipped at zero, D) fit by spline through points.}
    \label{fig:furnaceGraphs}
\end{figure}
In the lecture we explain the scenario and then start to break it into its component parts. We first discuss the data, and then how the data could be visualised. 

\begin{figure}
    \centering
    \includegraphics[alt={Text of the scenario focusing on who, what, why, how, when, who.},width=\linewidth]{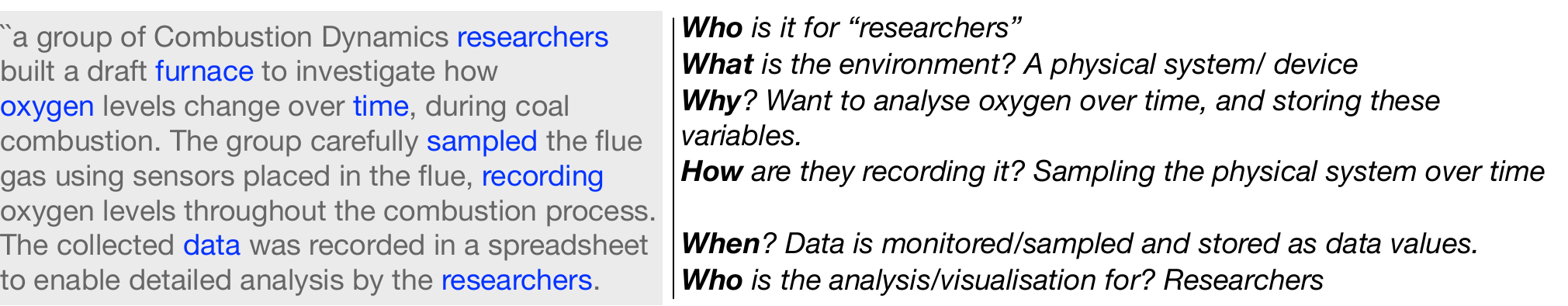}
    \caption{By focusing on the components of the system we can discuss different aspects and ramifications of the data and system to hand.}
    \label{fig:furnace-text}
\end{figure}

\subsection{Discussing the scenario, data and possible visualisations}
After introducing the scenario, \textbf{we guide students through a discussion of the system's components}, such as who the data is for, what’s being measured, and why it matters (see \cref{fig:furnace-text}). We also explore the different ways data can be collected: it might be simulated, observed manually, or gathered automatically. In this example, we assume the data was manually observed by researchers. We can discuss alternative methods, but for now we imagine the scientists using a separate device to measure oxygen levels, reading off values and recording them manually. This opens up opportunities to consider how data collection methods affect the accuracy, timing, and reliability of the data set;  from considering the sensor device, its accuracy, the method of storing the data and how humans make mistakes and so on.

After thinking through how the data is captured and stored, we move to how it could be visualised. In this scenario, there are challenges with how the data is visualised, especially without careful consideration, as shown in \cref{fig:furnaceGraphs}. Plotting the sampled oxygen levels with straight-line segments between each point (B) implies abrupt changes that do not reflect the continuous nature of the physical process. Alternatively, fitting a smooth polynomial curve (C) may result in the curve dipping below the x-axis, falsely suggesting negative oxygen levels, which are physically impossible in this context. Clipping the curve to the x-axis (D) avoids this issue visually but fails to accurately represent the underlying behaviour of the system. A more appropriate solution is to fit a constrained spline that ensures the curve remains above zero, better preserving both the physical reality and the integrity of the data.

This opens up a series of critical questions for students: Should the scientist have sampled more frequently? Is it feasible to rerun the experiment to fill in the missing values? Could simulation be used to estimate the missing points? Or is it better to fit the data with a curve that is constrained to respect domain-specific knowledge, such as the fact that oxygen levels cannot be negative?

This example represents a powerful case for encouraging critical thinking. Rather than simply plotting the data, students are asked to examine the broader context: How was the data sampled? What assumptions are made when fitting a curve? What happens when a high-order polynomial fit dips below zero, implying negative oxygen levels in the combustion system. Something physically impossible!
This scenario pushes students to go beyond surface-level visualisation. It asks them to put on their critical thinking hats and reflect on how they are mapping the data, what assumptions they are making, and what the final output communicates. Just as we use black-box testing or scenario-based evaluation in software development to ensure correctness, we should apply a similar mindset in data visualisation. We need to interrogate the structure of the data, the model it represents, and whether the visualisation faithfully reflects that model.

This scenario serves as a rich teaching vehicle. It encourages students to think critically about data modelling and visualisation choices. It highlights that creating a visualisation is not simply about plotting the data and moving on—it’s about understanding the system, interpreting the data responsibly, and communicating insights truthfully. The way the data is visualised, especially in cases like this where a naïve approach introduces error can significantly affect how results are perceived, and whether they are misinterpreted. It underlines the importance of thoughtful, informed visualisation practices in scientific analysis.

\section{Scenario 2: corridor footfall monitoring}
\label{sec:scenario_B}

\begin{figure}[h]
\begin{mdframed}
    
\begin{minipage}[t]{1.0\columnwidth}
\vspace*{0pt}
\textbf{Footfall monitoring.} 
After leaving university, Jack joined an IoT company and was hired by his former university to design and install a sensor system that automatically counts how many people walk down a corridor. He uses a microcomputer to store the data, which is then automatically sent to a remote database. The data is later filtered and visualised by the building's manager.
\end{minipage}

\vspace{5mm}
\begin{minipage}[t]{0.45\columnwidth}
\vspace*{0pt}
\centering
\begin{tabular}{|l|l|c|}
\hline
\textbf{Date} & \textbf{Time} & \textbf{Count} \\
\hline
2025-06-30 & 08:00–08:15 & 12 \\
2025-06-30 & 08:15–08:30 & 18 \\
2025-06-30 & 08:30–08:45 & 25 \\
2025-06-30 & 08:45–09:00 & 19 \\
2025-06-30 & 09:00–09:15 & 14 \\
2025-06-30 & 09:15–09:30 & 22 \\
\hline
\end{tabular}
\label{tab:footfall_data}
\end{minipage}
\hfill
\begin{minipage}[t]{0.3\columnwidth}
\vspace*{0pt}
\centering
\includegraphics[alt={Schematic diagram presenting the footfall monitor scenario.},width=0.9\columnwidth]{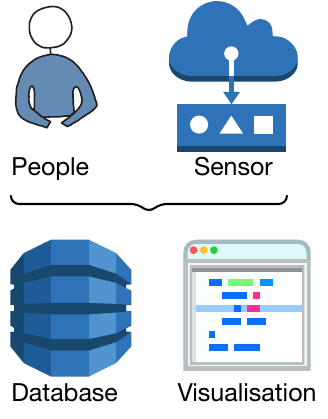}
\label{fig:footfall_monitor}
\end{minipage}
\end{mdframed}
\vspace{-5mm}
\caption{Scenario 2: Footfall Monitor. This scenario includes a written description of the data capture device and context, accompanied by a sample data table and an example diagram.}
\end{figure}

For the second scenario, we focus on footfall monitoring within a university building, \cref{fig:footfall_monitor}. The aim is to track how many people walk through a corridor, using a small sensor system installed down a corridor. The scenario gives a company, a developer and the environment where the device will be installed. This project reflects a common real-world application of data collection—monitoring physical spaces to support planning, safety, or resource management. The system automatically captures and stores data, which is later filtered and visualised by the building's manager. Through this scenario, we explore how sensor data is generated, stored, and interpreted, and consider the practical and ethical challenges involved in designing such a system.

In this second scenario, we explore a real-world example of footfall monitoring using Internet of Things (IoT) technology in a university corridor. The data is intended for the university's building manager, who uses it to monitor space usage and optimise operations such as heating, cleaning, and scheduling. The environment is a public corridor within a university building, and the data is recorded using a motion sensor system installed overhead. This raises important design considerations: what type of sensor is used-infrared, ultrasonic, or camera-based, and what resolution it operates at. Can it distinguish individuals, or detect anomalies such as people walking side-by-side, carrying bags, or using mobility aids like wheelchairs? Since this is real-world data, it's also vulnerable to manipulation as people could intentionally walk back and forth to inflate the count.

The sensor placement is also important: it's installed on the ceiling within a protective enclosure, but is that enclosure secure? Could someone tamper with it? The data is stored automatically in a remote database, transmitted via a networked microcomputer. However, this opens up additional questions: is the device dynamically connected to the network, or does the building manager need to manually access and download the data? What are the security implications of that connection? Furthermore, how often is the data recorded, is it per event, or batched in time intervals? And does the device perform any processing, such as averaging or filtering, before sending the data? Finally, the building manager is the end-user of this system, so we must ask: what are their technical skills? Are they capable of performing detailed analytics, or do they need simple, visual summaries? These questions help frame not only the scenario's technical considerations, but also the broader implications of designing visualisation tools that are useful, ethical, and user-appropriate.

\section{Abstracting the components of the scenario}
\begin{figure}
    \centering
    \includegraphics[alt={Overview of the scenario, with title, description, data table, and visualisation.},width=1\linewidth]{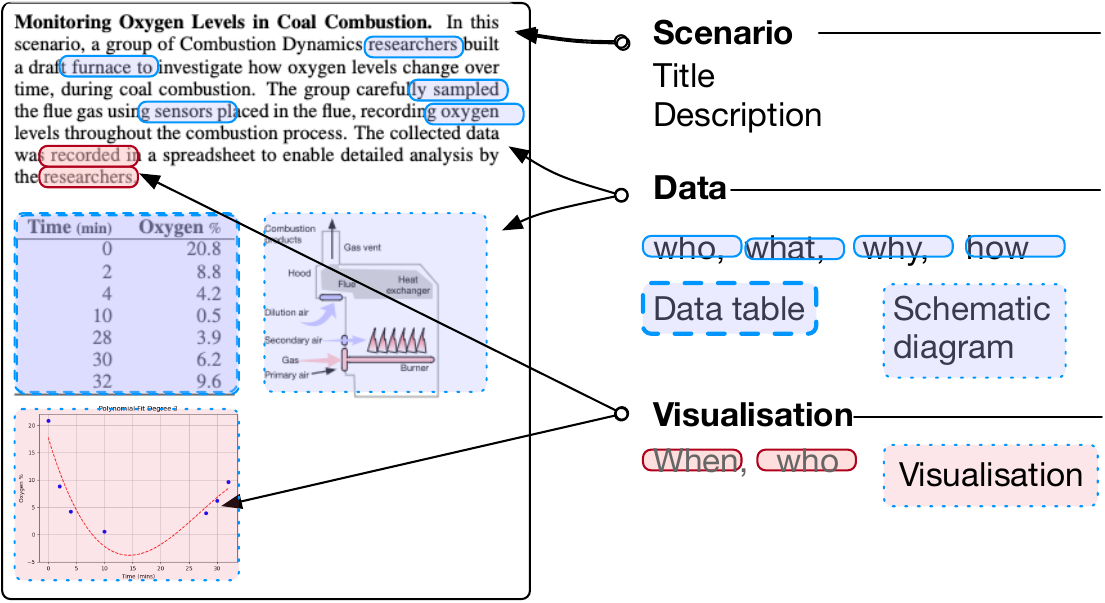}
    \caption{Overview of scenario components: (1) a scenario title and narrative description to establish context, (2) a breakdown of the data, including who collected it, why, where, when, and how, and (3) sample visualisations, considering appropriate forms and intended audience.}
    \label{fig:scenarioComponent}
\end{figure}

Each scenario follows a consistent structure, organised into two key stages, first, how the data is captured, and second, how it can be visualised, as depicted in \cref{fig:scenarioComponent}.  Every scenario begins with a title, which serves as a convenient reference point throughout the course. For instance, when we later mention the ``footfall monitoring example'', students immediately recall the context. The scenario also includes a concise description that outlines the purpose and scope; who the data is for, what they are trying to achieve, why the data is needed, and how the data is collected. This description should be informative yet open enough to allow for discussion and interpretation. The descriptions are concise, follow a consistent structure, and intentionally include cues that address the who, what, why, how, and when. They are crafted to be focused, engaging, and open-ended enough to encourage discussion and interpretation.

Next, each example includes a sample dataset, typically presented in a small table, and is  accompanied by a diagram. This might be a schematic illustration or a simplified visual showing the data capture process or the layout of the physical system. These visuals help students understand the structure of the system and how data flows through it.

In the second stage, the focus shifts to the affordances of the data. Which discusses how it can be visualised and interpreted? This includes considering the intended audience, the goals of the visualisation, and which aspects of the data should be highlighted. We prompt students to reflect on questions such as: Who needs to see this? What should be communicated? What visual form is most appropriate for the context? This stage encourages critical thinking about the design choices and communication strategies behind effective data visualisation.

Next, students explore a range of visualisation alternatives and critically interrogate each one. For example, in the oxygen combustion scenario we display several different graphs—straight-line fits, naïve polynomials, clipped curves, and constrained splines, and ask: Which version best represents the data? Is it accurate? Why or why not? We prompt them to assess suitability for the end user: Does this chart include all necessary elements (title, axes labels, legends)? Does it reveal or obscure key patterns? Are any data points misrepresented or omitted? Through guided questioning what's missing? How might you improve it?  Students learn to identify flaws, propose enhancements, and justify their design decisions, developing the habit of rigorous critique before finalising any visualisation.

We also discuss how data can be classified by measurement scale; nominal (categorical labels), ordinal (ranked categories), interval (numerical values without a true zero), and ratio (numerical values with a meaningful zero), and how these distinctions guide our choice of visual encodings. Equally important is understanding the origin of the data: observational data (real-time sensor readings, surveys, images) are often unique and irreversible; experimental data (lab measurements such as gene sequences or spectroscopy) can be replicated but at greater cost and logistical effort; simulation data (climate or economic models) depend as much on their underlying assumptions and metadata as on their outputs; derived or compiled data (aggregated from multiple sources or mined through algorithms) require careful provenance tracking; and reference or canonical data (peer-reviewed databanks like census records or gene repositories) offer high reliability but demand attention to licensing and versioning. By covering both scale and origin in a single discussion, we help students appreciate how the nature of a dataset influences every stage of its visualisation, from cleaning and transformation to chart selection and annotation.

\section{Discussion}
These data scenarios go beyond simple teaching aids, providing essential context and acting as strong narrative hooks that engage students with both the technical and conceptual challenges of visualisation. They help educators move past chart types and graphical techniques, fostering richer discussions about ethics, accuracy, and interpretation in data-driven work, while also encouraging students to reflect on their own data. The contextual scenarios outlined in this paper are incorporated into student assessments, and the critical thinking exercises demonstrated in lectures help students embed these concepts into their work. In the visualisation module, students select a dataset and conduct an analysis following a process similar to the lectures. They then produce a data report and a design study, using the five design-sheets method \cite{RobertsHeadleandRitsos15_TVCG} alongside the Critical Design Strategy (CDS) \cite{RobertsAlnjarOwenRitsos2026} to support critical thinking. For their second assessment, they create a visualisation of their dataset and submit a critical report reflecting on their design choices and outcomes.

By organising each scenario into two distinct phases, first, data capture (who, what, where, when, how) and then visualisation affordances (intended audience, objectives, and design decisions), we establish a clear, scaffolded framework for learning. This structure transforms lectures from an unconnected series of concepts into a cohesive sequence of steps. It also allows instructors to rapidly develop new examples, keep course materials neatly organised, and easily reference scenarios by their concise titles in future discussions.

We started collecting feedback from students. We asked students to provide feedback on their experience in the information visualisation module. The end-of-module questionnaires were voluntary and were achieved online and anonymously. We received a rating of 93\% satisfaction for the whole teaching, lecturing and so on, which is a positive outcome. In previous years, students said that they'd struggled with lectures but found the class activities useful, whereas with this approach we received comments such as ``Both the lectures and practical sessions were valuable to me. None was more valuable than the other''. We asked students to write about their experiences and without specifically asking for an answer about the data scenarios, a student wrote ``The most appreciated aspects include: Practical Case Studies: Real-world scenarios that enhanced understanding and applicability'', and another wrote ``the intellectually stimulating content was the most valuable, as it challenged me to think critically about visualisation techniques and applications''. This is positive qualitative feedback and is supportive of the example based lectures. While a more in-depth study is required, it presents a positive and encouraging outcome.

The two scenarios presented here represent the starting point of a larger, ongoing body of work. They were selected because they illustrate physical systems in which measurement errors and uncertainty play a significant role. For example, in the footfall monitoring case, the presence of a trolley bag can alter sensor readings. Such situations prompt students to grapple with the practical challenges of data reliability and interpretation. Beyond these examples, we have developed a wider set of scenarios that address additional considerations, including societal relevance, demographics, and equity-related questions. In addition, students explore these issues alongside project work, which forms part of an aligned Research Methods module. In that course, students engage explicitly with ethics, human bias, and representation, and are required to address these issues within their project submissions. The present paper focuses on two illustrative cases to demonstrate our approach, while a more detailed treatment of the broader scenario set is left to future work.

A frequent challenge with authentic assessments, and with scenarios like those described here, is that they may not exactly match the specific tasks or contexts students will face in their professional lives. Given the diversity of professional pathways and evolving nature of data-related roles, it is inherently difficult to predict the exact challenges students will face after graduation. However, the strength of these scenarios lies in their ability to expose students to a broad spectrum of scenarios and practical issues, thus cultivating adaptable problem solving skills. By engaging in a variety of data contexts and associated challenges, students develop a flexible mindset and critical awareness that can be transferred to unfamiliar or novel situations. This broad exposure encourages students to consider diverse practical concerns, ranging from technical errors to ethical implications, equipping them with a toolkit that transcends any single domain or task.

Finally, embedding these scenarios within an authentic learning framework encourages students to take ownership of their work. Early ``data triage'' exercises teach them to assess dataset suitability and ethical considerations, mirroring professional practices. Personalisation allows learners to pursue topics aligned with their interests while still applying the same rigorous scenario structure. Regular formative feedback and real-world relevance, whether it's footfall monitoring in a university corridor or oxygen‐level measurements in a combustion experiment ensure that students not only learn visualisation techniques but also internalise the critical thinking, contextual awareness, and metacognitive skills essential for responsible, impactful data storytelling.


\acknowledgments{
The authors wish to thank the many students who have given feedback on the lectures.}

\bibliographystyle{abbrv-doi-hyperref}
\balance

\bibliography{context-scenarios-new}
\end{document}